\documentclass{ws-procs9x6}

\usepackage[center,footnotesize,hang]{subfigure}

\allowdisplaybreaks[1] 
 
\newcommand{\CenterEps}[2][1]{
\ensuremath{\vcenter{\hbox{\includegraphics[scale=#1]{#2.eps}}}}
}

\newcommand{\RaiseBrace}[1]{\raise3pt\hbox{$\displaystyle#1$}}

\def\<{\left\langle}
\def\>{\right\rangle}

\begin{document}

\title{Implications of Running Neutrino Parameters 
for Leptogenesis 
and for \\ Testing Model Predictions
}

\author{\vspace{-2mm}Stefan Antusch
}

\address{
Department of Physics and Astronomy,
University of Southampton,\\
Southampton, SO17 1BJ, United Kingdom, \\
E-mail: santusch@hep.phys.soton.ac.uk}

\maketitle

\abstracts{
The running of neutrino parameters in see-saw models and its 
implications for leptogenesis 
and for testing predictions 
of mass models with future precision experiments are discussed 
using analytical
approximations as well as numerical results. 
}

\vspace{-2mm}
\section{Introduction}
Fermion mass models and the leptogenesis\cite{Fukugita:1986hr} mechanism for explaining the baryon
asymmetry of our universe typically operate at high energies  
close to the unification scale $M_\mathrm{U}$ or at the see-saw scales, 
i.e.\ the masses 
the heavy right-handed neutrinos. Our knowledge about the neutrino masses and
mixings on the other hand mainly stems from experiments on neutrino
oscillations, performed at low energy. In order to compare the high-energy
predictions with the low energy experimental data, the renormalization group
(RG) running of the relevant quantities has to be taken into account. In 
see-saw models for neutrino masses (type I and type II), this requires solving the 
RGEs\cite{b1}\cdash\cite{Grimus:2004yh} for the effective 
neutrino mass matrix for the various effective theories which arise from 
successively integrating out the heavy degrees of freedom, in particular the 
heavy right-handed neutrinos.

\vspace{-2mm}
\section{Implications for Leptogenesis}
For the leptogenesis mechanism, the relevant scale is 
the mass $M_1$ of the
lightest right-handed neutrino, or, in the type II case, possibly also the mass
scale $M_\Delta$ of the lightest SU(2)$_\mathrm{L}$-triplet. In the energy range between the leptogenesis
scale and the electroweak scale $M_\mathrm{EW}$, we can consider the running of
the effective neutrino mass operator, which is produced from integrating out the
heavy right-handed neutrinos and/or triplets.

\vspace{-2mm}
\subsection{Enhancement of the Decay Asymmetries}\label{sec:EnhandedEps}
\begin{figure}
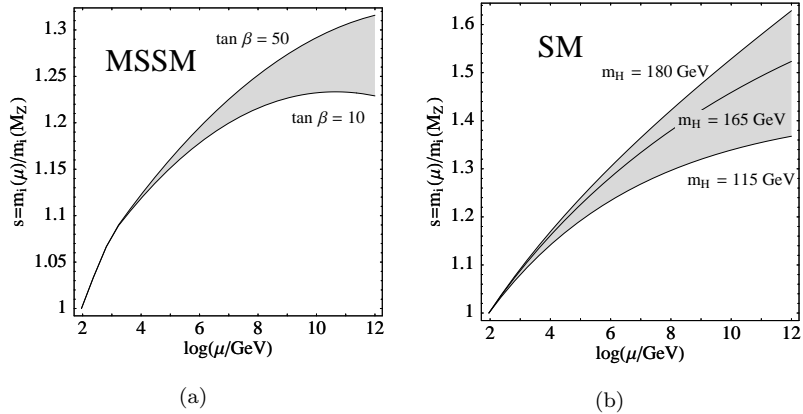

\begin{center}
     \subfigure[]
         {$\CenterEps[0.595]{Scaling_MSSM_178}$} 
    \hfil
     \subfigure[]
         {$\CenterEps[0.595]{Scaling_SM_178}$} 
\end{center}
\vspace*{-7mm}
 \caption{\label{fig:RGscalingMnu}
 Scaling of the neutrino mass eigenvalues $m_i$ from the 
 RG evolution of the effective neutrino mass matrix in the MSSM (Fig.\ 1(a)) 
 and in the SM (Fig.\ 1(b)).$^{12}$   
 In the MSSM with large $\tan \beta$, we have shown the evolution of 
 $m_3$ for a normal mass
 ordering and with CP phases set to $0$. 
 For the plots, $m_t = 178\:$GeV has been used. 
   } 
\end{figure}

The decay asymmetries\cite{Covi:1996wh,Antusch:2004xy,Hambye:2003ka} 
for type I leptogenesis 
as well as for type II 
leptogenesis 
(via the 
lightest right-handed neutrino) can be written as 
$\varepsilon_1 \sim 
- \frac{M_1}{v_\mathrm{EW}^2} \<m^{\mathrm{BAU}}_{\mathrm{eff}}\>$, where 
$\<m^{\mathrm{BAU}}_{\mathrm{eff}}\> := 
\frac{1}{(Y_\nu^\dagger Y_\nu)_{11}} 
\sum_{fg}\mbox{Im}\, [(Y^*_\nu)_{f1} (Y^*_\nu)_{g1} 
 (m{}_{\nu})_{fg}]$ is an effective mass for leptogenesis.  
$Y_\nu$ is the neutrino Yukawa matrix and we have considered the case of 
hierarchical right-handed neutrino masses and $M_\Delta \gg M_1$. 
In the SM or for a moderate
$\tan \beta$ in the MSSM, the RG running from high to low energy 
leads mainly to a scaling of the neutrino mass matrix $m_\nu$  
(see Fig.~\ref{fig:RGscalingMnu}). 
Including the RG effects thus leads to an 
enhancement of the decay 
asymmetry for leptogenesis\cite{Barbieri:1999ma,Antusch:2003kp} by a factor of roughly 
20\% in the MSSM and 30\% - 50\% in the SM.\cite{Antusch:2003kp}

\vspace{-4mm}
\vspace{-4mm}

\vspace{-2mm}
\subsection{Correction to the Bound on the Neutrino Mass Scale}
From the requirement of successful thermal type I leptogenesis,  
a bound on the absolute neutrino
mass scale can be derived.\cite{Buchmuller:2003gz}  
Among the significant corrections to this bound, 
included in recent calculations, are the effects from RG running between low energy 
and $M_1$. 
With an increased 
decay asymmetry $\varepsilon_1$ as discussed in section \ref{sec:EnhandedEps}, 
the produced baryon asymmetry 
increases as well.
However, scattering precesses which tend to wash out the baryon
asymmetry are enhanced as well, which typically   
over-compensates the correction
to the bound from
the enhanced decay asymmetry 
and makes the bound on the neutrino mass 
scale more
restrictive.\cite{Antusch:2003kp,Giudice:2003jh,Buchmuller:2004nz}

\vspace{-2mm}
\subsection{Implications for Resonant Leptogenesis}         
Resonant leptogenesis\cite{PilaftsisPascos04} relies on a small splitting between the masses of the
lightest right-handed neutrinos, $M_1$ and $M_2$, of the order of their decay
widths. Given a model for neutrino
masses with such a small mass splitting defined at $M_\mathrm{U}$, it can be
affected significantly by the RG evolution of the mass matrix of the heavy
right-handed neutrinos from
$M_\mathrm{U}$ to $M_1 \approx M_2$. On the other hand, one can also 
have exactly degenerate masses $M_1 = M_2$ at high energy and 
generate the required mass splitting radiatively.\cite{GonzalezFelipe:2003fi}

\vspace{-2mm}
\section{Implications for Testing Model Predictions by Future Precision
Experiments}
Future reactor and long-baseline experiments have the potential to measure 
the neutrino mixing angles to a high precision. For testing the predictions of 
 mass models
  using such precision measurements, the RG corrections will be
 important, in particular, if the observed mixing angles turn out to be close to
 theoretically especially interesting values. 
 For conservative estimates of the RG effects, analytical 
 formulae\cite{Chankowski:1999xc,Casas:1999tg,Antusch:2003kp} for the
 running of the neutrino parameters below the see-saw scales 
 can be used.  
 For an accurate determination of
 the RG effects in specific models, the model dependent 
 running above and between the see-saw scales can 
contribute significantly and often even dominates the RG
effects.\cite{King:2000hk,SeesawRunning}  
Formulae which allow an analytic understanding of the running above 
the see-saw scales are in preparation.\cite{preparation} 

\vspace{-2mm}
\subsection{Radiative Generation of $\theta_{13}$}
One important parameter is the mixing angle
$\theta_{13}$. The knowledge of its
value will allow to discriminate between many fermion mass models and 
furthermore, only if $\theta_{13}$ is not too small, future experiments on 
neutrino oscillations have the potential to measure leptonic CP violation.  
Do we expect $\theta_{13}$ very close to zero at low energy? 
Even if $\theta_{13}=0$ 
 is predicted by some model at high energy, 
 RG running will in general generate $\theta_{13}\not=0$ at low energy. 
From a conservative estimate using the
analytical formulae below the see-saw scales, it has been shown\cite{Antusch:2003kp} 
that the RG corrections are often comparable to, or even exceed the expected sensitivity of future 
experiments. 
Note that for $\theta_{13}$,  
 small values of CP phases (as predicted e.g.\ 
by certain type II see-saw models\cite{Antusch:2004xd})   
can protect against large RG corrections.\cite{Antusch:2003kp} 
Radiative generation of $\theta_{13}$ from running  
above the see-saw scales has been analyzed in Ref.~\refcite{Mei:2004rn}.     

\vspace{-2mm}
\subsection{Modification of Complementarity Relations for $\theta_{12}$}
With the present neutrino data, complementarity 
relations\cite{Petcov:1993rk}\cdash\cite{Frampton:2004vw} such as  
$\theta_{12} + \theta_C = \pi/4$ (with $\theta_C$ being the Cabibbo angle) 
are allowed and will be tested by future 
experiments to a high accuracy. 
RG corrections can lead to significant modifications of such 
relations for $\theta_{12}$.\cite{Minakata:2004xt}

\vspace{-2mm}
\subsection{Corrections to Maximal Mixing $\theta_{23}$} 
The present best-fit value for $\theta_{23}$ is close to maximal. Typically,
mass models predict a deviation of $\theta_{23}$ from maximality,  
which is within reach of future long baseline experiments.\cite{Antusch:2004yx}
If $\theta_{23}$
turns out to be close to maximal, this would point towards a symmetry which
fixes maximal mixing at high energy $M_\mathrm{U}$. 
However, even if $\theta_{23}=\pi/4$ is predicted by some model at high energy,  
RG corrections
from the running between $M_\mathrm{U}$ and low energy generate a deviation
of the low energy value for $\theta_{23}$ from 
maximality.\cite{Antusch:2003kp,Antusch:2004yx} In many cases, 
even for hierarchical neutrino masses, this deviation is comparable
to, or exceeds the sensitivity of future 
experiments (see Fig.\ 2).       

\vspace{-5mm}
\begin{figure}
\begin{center}
     \subfigure[Experimental sensitivity]
         {$\CenterEps[1]{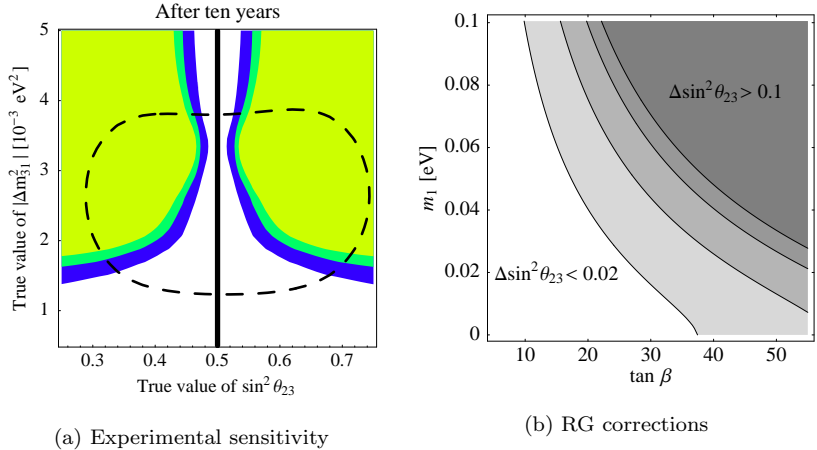}$}  
    \hfil
     \subfigure[RG corrections]
         {$\CenterEps[0.635]{RGcorrectionsT23_1}$} 
\end{center}
\vspace*{-5mm}
 \caption{
 Fig.~2(a) shows the expected sensitivity of future long-baseline experiments 
 (combined MINOS, ICARUS, OPERA, JPARC-SK and MuMI; 
 see Ref.~28 for details) for excluding maximal mixing 
 $\sin^2 \theta_{23}=0.5$ in 10 years at $1\sigma$, $2\sigma$ and
 $3\sigma$ (from light to dark shading). 
 The dashed line shows the currently allowed region for $\theta_{23}$ at $3\sigma$. 
 Fig.~2(b) (from Ref.~28; see also Ref.~12 for details) shows a
 conservative estimate (ignoring $Y_\nu$-effects) for the RG corrections 
 to maximal $\theta_{23}$ from 
 the running between $M_\mathrm{U}$ and $M_\mathrm{EW}$ in the MSSM.  
 The contour lines correspond to 
 $\Delta\sin^2 \theta_{23}=0.02$, $0.05$, $0.08$ and $0.1$. 
 } 
 \label{fig:ScalingMnu}
\end{figure}

\vspace{-2mm}
\section{Summary and Conclusions}
Facing the high expected sensitivities of future experiments on the neutrino
parameters, RG corrections are increasingly relevant for testing  
predictions of mass models. They are particularly important, if the 
neutrino mass spectrum turns out to be non-hierarchical or if 
experiments find the lepton  
mixing angles close to specific values such as $0$ for $\theta_{13}$,
$\pi/4$ for $\theta_{23}$ or compatible with 
complementarity relations such as $\theta_{12} + \theta_C = 
\pi/4$ (with $\theta_C$ being the Cabibbo angle) for $\theta_{12}$.  
For leptogenesis, the scaling of the neutrino masses by RG effects enhances the
decay asymmetries for type I/II leptogenesis, effects washout parameters and
finally lowers the bound on the
absolute neutrino mass scale from the requirement of 
successful thermal type I leptogenesis.

%

\end{document}